\documentclass[aps,pre,superscriptaddress,twocolumn]{revtex4-1}
\usepackage{bm}
\usepackage{graphicx}
\usepackage{tikz}
\usepackage{amsmath}
\usepackage{float}
\usepackage{amsfonts}
\usepackage{graphicx}
\usepackage{xcolor}
\usepackage{epstopdf}
\usepackage{booktabs}

\begin{document}

\title{Renormalization of stochastic differential equations with multiplicative noise using effective potential methods}

\author{Jean-S\'{e}bastien Gagnon}
\email[]{jgagnon6@norwich.edu}
\affiliation{Physics Department, Norwich University, Northfield, Vermont, USA}
\affiliation{Department of Earth and Planetary Sciences, Harvard University, Cambridge, Massachusetts, USA}

\author{David Hochberg}
\email[]{hochbergd@cab.inta-csic.es}
\affiliation{Department of Molecular Evolution, Centro de
Astrobiolog\'{i}a (CSIC-INTA), Torrej\'{o}n de Ard\'{o}z, Madrid,
Spain}

\author{Juan P\'{e}rez-Mercader}
\email[]{jperezmercader@fas.harvard.edu}
\affiliation{Department of Earth and Planetary Sciences, Harvard University, Cambridge, Massachusetts, USA}
\affiliation{Santa Fe Institute, Santa Fe, New Mexico, USA}

\begin{abstract}
We present a new method to renormalize stochastic differential equations subjected to multiplicative noise.  The method is based on the widely used concept of effective potential in high energy physics, and has already been successfully applied to the renormalization of stochastic differential equations subjected to additive noise.  We derive a general formula for the  one-loop effective potential of  a single  ordinary stochastic differential equation (with arbitrary interaction terms) subjected to multiplicative Gaussian noise (provided the noise satisfies a certain normalization condition).   To illustrate the usefulness (and limitations) of the method, we use the effective potential to renormalize a toy chemical model based on a simplified Gray-Scott reaction.  In particular, we use it to compute the scale dependence of the toy model's parameters (in perturbation theory) when subjected to a Gaussian power-law noise with short time correlations. 
\end{abstract}

\date{\today}

\maketitle

\section{Introduction}
\label{sec:Introduction}

A variety of systems can be modeled using stochastic differential equations (SDEs).
Examples include population dynamics (e.g.~\cite{Dobramysl_etal_2018}), surface growth (e.g.~\cite{Kardar_etal_1986,Barabasi_Stanley_1995}),
pattern formation (e.g.~\cite{Lesmes_etal_2003,Karig_etal_2018}), and financial markets (e.g.~\cite{Black_Scholes_1973}), to name a
few (more applications can be found in Refs.~\cite{Garcia_Sancho_1999,Horsthemke_Lefever_2006,vanKampen_2007}).  In the above examples, the
noise term is often introduced by hand to take into account the effects of unaccounted degrees of freedom on the system (such as the environment), or
appears naturally as a result of intrinsic stochasticity (due to effects of finite number of constituents, for example).

In addition to SDEs, there exists other approaches to study theoretically the effect of noise on a system, all with their pros and cons.  For example, the Master equation~\cite{vanKampen_2007,Gardiner_2004} is unwieldy, but it is well-suited to study intrinsic noise, i.e. noise that is due to the discrete nature of the system and cannot be shut off.  By contrast, the degrees of freedom in SDEs are assumed to be well-defined ``macroscopic'' quantities (e.g. concentrations).  Consequently, approaches based on SDEs do not take into account intrinsic noise at a fundamental level ---although intrinsic noise can be obtained from a more fundamental approach, and then added by hand to the SDE (see Ref.~\cite{Karzazi_etal_1996} for an explicit example related to reaction-diffusion systems).  


Our own interest lies in chemical systems {\em externally} driven with noise, where the noise can either be viewed as a control tool (that could be used to implement chemical logic gates~\cite{Egbert_etal_2018,Egbert_etal_2019} or
force chemical systems into dynamical regimes not otherwise easily attainable~\cite{Srivastava_etal_2018}), as a probe of chemical mechanisms~\cite{Gagnon_PerezMercader_2017}, or as an environment that
influences the behavior of the chemical system (with origins of life applications in mind,  such as the appearance of homochirality in biomolecules~\cite{Hochberg_2010}).  In the above applications, the noise is external and the number of constituents is sufficiently large for a description in terms of SDEs to be appropriate.  We concentrate on such systems in the following.

Common approaches to the study of noise in physical systems (SDEs, Master equation, Fokker-Planck equation, etc) focus on determining the time evolution of the statistical properties of the solution directly.  This can be a difficult task, and not always necessary for all applications.  For instance, we show in Ref.~\cite{Gagnon_PerezMercader_2017} how measuring variations in a chemical system's parameters (e.g. reaction rates) due to tunable external noise can be used as a probe of chemical mechanisms.  This is akin to the idea in particle physics of measuring variations in a system's parameters (coupling constants, charges, etc) with great precision in order to uncover details about the underlying (high energy) theory.  

It is the complex interplay between fluctuations and
interactions that leads to scale-dependence in the parameters and
couplings in stochastic systems, where the fluctuations can be of
either thermal or statistical nature. This scale-dependence can be 
unveiled by the renormalization group, whose aim is to describe how
the dynamics of the system evolves as we change the temporal and or
spatial scale at which the phenomenon is being observed or
measured.

Two broad categories of noise ---additive and multiplicative--- are particularly relevant for physical applications.  Noise in SDEs is considered additive when it is added to terms containing the dynamical variables.  The original Langevin equation describing Brownian motion and the Kardar-Parisi-Zhang equation~\cite{Kardar_etal_1986} are examples of equations with additive noise.  In contrast, multiplicative noise is operative when the noise term multiplies the dynamical variables, i.e. the effect of the noise depends on the state of the system.  Examples include the use of diffusion equations with random sources and sinks to model directed polymers in random media~\cite{Kardar_Zhang_1987} and stochastic systems with an absorbing state~\cite{Munoz_1998}.

In Refs.~\cite{Hochberg_etal_2003,Gagnon_etal_2015,Gagnon_PerezMercader_2017,Gagnon_etal_2017,Gagnon_etal_2018}, perturbative renormalization group methods~\cite{Medina_etal_1989,Barabasi_Stanley_1995} are used to compute the running of parameters for additive power-law noise in a simple cubic autocatalytic reaction-diffusion model. In this toy model, the additive noise represents possible fluctuations in the inflow of chemicals into the system, and constitutes a possible way for an experimentalist to control the dynamic of the reaction.  Another interesting ``control knob'' is to use noisy light to influence the rate of light-sensitive reactions~\cite{Simakov_PerezMercader_2013,Simakov_PerezMercader_2013b,Munuzuri_PerezMercader_2017}.  Since the fluctuating reaction rate now multiplies one of the degrees of freedom of the system (i.e. concentration), the noise is multiplicative.   However, applying the same methods developed in Refs.~\cite{Hochberg_etal_2003,Gagnon_etal_2015,Gagnon_PerezMercader_2017,Gagnon_etal_2017,Gagnon_etal_2018} to SDEs subjected to multiplicative noise is untenable, as there is no simple way of truncating the formal solution to the differential equation. To our knowledge, and despite progress in analytical~\cite{Schenzle_Brand_1979_1,Schenzle_Brand_1979_2,Brey_etal_1987,Denisov_Horsthemke_2002,Moreno_etal_2019} and numerical~\cite{Garcia_Sancho_1999,Sato_etal_2000,Dornic_etal_2005,Cassol_etal_2012} techniques, the computation of running parameters for stochastic differential equations subject to multiplicative noise is still lacking (a notable exception is Ref.~\cite{Schoner_1985}, where the author uses the Martin-Siggia-Rose formalism to renormalize two-point correlation functions in real time).


In this paper, we propose a new method to renormalize stochastic
differential equations subjected to multiplicative noise based on the physical concept of the effective potential.  In the case of quantum field theory, the eminent role of the effective action, and its specialization to constant fields, the effective potential, as fundamental constructs for obtaining this scale-dependence has been recognized and exploited for a long time \cite{Goldstone_etal_1962,JonaLasinio_1964,Fujimoto_etal_1983,Gato_etal_1985,Hochberg_etal_1999b,AlvarezGaume_2012}.  Yet, it is only in the last $20$ years that the concepts of effective action and
effective potential have been defined, calculated and applied
successfully to quantify the scale-dependence for stochastic systems
subject to additive noise~\cite{Hochberg_etal_1999}.  

We here assume from the outset continuous and differentiable degrees of freedom in order that the powerful field concept can be applied directly to the study of
fluctuation phenomena.  Starting from a rather general class of nonlinear SDEs, an explicit construction was introduced in Ref.~\cite{Hochberg_etal_1999} which
maps the SDE into an associated characteristic functional.  This characteristic functional is then used to define and calculate the effective action and the effective potential.
There it was demonstrated that the potential can be used to quantitatively assess the impact that noise and random fluctuations can have in altering the ``ground'' states of
dynamic fluctuating systems.  By borrowing  fundamental concepts from quantum field theory, one can, by means of mapping of the SDEs to its characteristic generating functional, set up
and calculate the effective action and its specialization to static field, which yields the effective potential.  As discussed in
Ref.~\cite{Hochberg_etal_1999}, if some notion of a potential is available, then the analysis of its extrema leads to the identification of the stable and metastable states
of the fluctuating system.  These ideas carry over to non-equilibrium systems and the effective potential constructed from Langevin equations contains two
major pieces: a ``classical'' contribution and a ``fluctuating'' contribution.  Both contributions together determine the ground state of the system in the presence of noise and
therefore the effective potential provides useful information and gives one another way to understand out-of-equilibrium fluctuation phenomena from a distinct vantage point.
These ideas and concepts have been applied to the Kardar-Parisi-Zhang equation (for which dynamical symmetry breaking phenomena results in one and two dimensions for white
Gaussian noise)~\cite{Hochberg_etal_2000,Hochberg_etal_2001} and reaction-diffusion equations~\cite{Hochberg_etal_2000b}.

The  primary  goal of this paper is to generalize the effective potential method of Ref.~\cite{Hochberg_etal_1999} to
multiplicative noise.  The expression for the effective potential obtained here is applicable to a single ordinary differential equation (with arbitrary interaction terms) subjected to multiplicative Gaussian noise, provided the noise satisfies a certain normalization condition.  To illustrate the usefulness (and limitations) of the effective potential in the context of SDEs, we use it to renormalize a toy chemical model based on a simplified Gray-Scott reaction subjected to multiplicative noise.  In particular, we show that for a Gaussian power-law noise with short time correlations, the removal (or decay) rate of the reaction varies with scale (provided perturbation theory is valid).

The rest of this paper is organized as follows.  We first show in Sect.~\ref{sec:Perturbative_RG} the difficulty in applying
conventional perturbative renormalization group (RG) techniques to SDEs subjected to multiplicative noise.  We then derive the
main formula for the effective potential generalized to multiplicative noise in Sect.~\ref{sec:Effective_potential_multiplicative_noise}, and apply it to a
toy model example in Sect.~\ref{sec:Effective_potential_methods_toy_model}.

\section{Perturbative RG and multiplicative noise}
\label{sec:Perturbative_RG}

One motivation for introducing the effective potential method to renormalize SDEs subjected to multiplicative noise is that the more conventional perturbative RG method~\cite{Medina_etal_1989,Barabasi_Stanley_1995} seems to be ill-suited for the task.  To show this explicitly, consider the following additive noise SDE:
\begin{eqnarray}
\label{eq:Toy_model_additive_noise}
\frac{d\phi(t)}{dt} & = & -r\phi(t) + \lambda\phi^{2}(t) + \eta(t),
\end{eqnarray}
and a possible multiplicative noise variant:
\begin{eqnarray}
\label{eq:Toy_model_multiplicative_noise}
\frac{d\phi(t)}{dt} & = & -r\phi(t) + \lambda\phi^{2}(t) + \phi(t)\eta(t),
\end{eqnarray}
where $r$ and $\lambda$ are parameters and $\eta(t)$ is a noise term.   Equation~(\ref{eq:Toy_model_additive_noise}) represents a simplified version of the cubic autocatalytic chemical reaction model studied in Refs.~\cite{Gagnon_etal_2018}.  It involves a certain chemical $\phi$ reacting with another chemical U via the autocatalytic reaction $2\phi$ + U $\rightarrow$ $3\phi$, and the removal of $\phi$ from the continuously stirred tank reactor (this is a variant of the Gray-Scott reaction in which $U$ is considered very abundant and thus having a constant concentration).  This model can be seen as a crude form of metabolism, where the ``organism'' (represented by the chemical $\phi$) produces more of itself using the ``food'' U.  The noise term in Eq.~(\ref{eq:Toy_model_additive_noise}) could correspond to some external factor that influences the rate of production of $\phi$, while the noise term in Eq.~(\ref{eq:Toy_model_multiplicative_noise}) represents a fluctuating removal rate.   Note that in this paper, we adopt the Stratonovich interpretation for the noise (see Appendix 1 in Ref.~\cite{Hochberg_etal_1999} for explanations on this choice).

 To perform the perturbative RG analysis of the above toy models, we  insert the Fourier representation for the field and noise  (hats denote Fourier transformed quantities):
\begin{eqnarray}
\phi(t) & = & \int_{-\infty}^{\infty}\frac{d\omega}{(2\pi)}\; e^{-i\omega t}\hat{\phi}(\omega), \\
\eta(t) & = & \int_{-\infty}^{\infty}\frac{d\omega}{(2\pi)}\; e^{-i\omega t}\hat{\eta}(\omega),
\end{eqnarray}
into Eqs.~(\ref{eq:Toy_model_additive_noise})-(\ref{eq:Toy_model_multiplicative_noise}) to get:
\begin{equation}
\label{eq:Iterative_equation_field_additive_noise}
\hat{\phi}(\omega) = \hat{R}_{0}(\omega)\left[\hat{\eta}(\omega) + \lambda\int\frac{d\omega_{1}}{(2\pi)}\hat{\phi}(\omega_{1})\hat{\phi}(\omega-\omega_{1})\right],
\end{equation}
for the additive noise case and:
\begin{eqnarray}
\label{eq:Iterative_equation_field_multiplicative_noise}
\hat{\phi}(\omega) & = & \hat{R}_{0}(\omega)\left[\int\frac{d\omega_{1}}{(2\pi)} \hat{\phi}(\omega-\omega_{1})\hat{\eta}(\omega_{1}) \right. \nonumber \\
             &   & \left. + \lambda\int\frac{d\omega_{1}}{(2\pi)}\hat{\phi}(\omega_{1})\hat{\phi}(\omega-\omega_{1}) \right],
\end{eqnarray}
for the multiplicative noise case, where $\hat{R}_{0}(\omega) = 1/(-i\omega + r)$ is the free response function for both SDEs.  Note that here and in the following, all integration boundaries are from $-\infty$ to $+\infty$, unless otherwise stated.  Defining the full response function as $\hat{\phi}(\omega) \equiv \hat{R}(\omega)\hat{\eta}(\omega)$, we can re-write  Eqs.~(\ref{eq:Iterative_equation_field_additive_noise})-(\ref{eq:Iterative_equation_field_multiplicative_noise}) as:
\begin{eqnarray}
\label{eq:Iterative_equation_G_additive_noise}
\lefteqn{\hat{R}(\omega)\hat{\eta}(\omega)  =  \hat{R}_{0}(\omega)\hat{\eta}(\omega)}  \nonumber \\
                      &   & + \lambda \hat{R}_{0}(\omega)\int\frac{d\omega_{1}}{(2\pi)} \hat{R}(\omega_{1}) \hat{R}(\omega-\omega_{1}) \hat{\eta}(\omega_{1})\hat{\eta}(\omega-\omega_{1}),\;\;\;
\end{eqnarray}
for the additive noise case and:
\begin{eqnarray}
\label{eq:Iterative_equation_G_multiplicative_noise}
\lefteqn{\hat{R}(\omega)\hat{\eta}(\omega) =  \hat{R}_{0}(\omega)\int\frac{d\omega_{1}}{(2\pi)} \hat{R}(\omega-\omega_{1})\hat{\eta}(\omega-\omega_{1})\hat{\eta}(\omega_{1})}\;  \nonumber \\
                      &   &  + \lambda  \hat{R}_{0}(\omega)\int\frac{d\omega_{1}}{(2\pi)} \hat{R}(\omega_{1}) \hat{R}(\omega-\omega_{1}) \hat{\eta}(\omega_{1})\hat{\eta}(\omega-\omega_{1}),\;\; 
\end{eqnarray}
for the multiplicative noise case.  The iterative equations~(\ref{eq:Iterative_equation_G_additive_noise})-(\ref{eq:Iterative_equation_G_multiplicative_noise}) can be used as a starting point for a perturbative analysis of the full response function.  In order to implement a perturbative renormalization program, it is necessary to truncate the above infinite series at a finite order in some small parameter (typically involving the amplitude of the noise).  This is relatively easy to do in the additive noise case, due to the presence of a zeroth order term that does not depend on the full response function in Eq.~(\ref{eq:Iterative_equation_G_additive_noise}) (i.e. $\hat{R} = \hat{R}_{0} + O(\hat{R})$).  The absence of such zeroth order term in Eq.~(\ref{eq:Iterative_equation_G_multiplicative_noise}) makes the task of truncating the infinite series difficult (it at all possible) for the multiplicative noise case.  Thus a different approach is warranted.

\section{Effective potential methods for multiplicative noise}
\label{sec:Effective_potential_multiplicative_noise}

The effective potential for stochastic partial differential equations subjected to additive noise is derived in detail in Ref.~\cite{Hochberg_etal_1999}.  Here we generalize this result to multiplicative noise, and outline the main steps below.  For simplicity, we focus on ordinary differential equations, but spatial derivatives can be added with minimal efforts.
We are interested in SDEs of the form:
\begin{eqnarray}
\label{eq:General_SDE_multiplicative}
\frac{d\phi(t)}{dt} & = & -F(\phi) + G(\phi)\eta(t),
\end{eqnarray}
where $F(\phi)$ and $G(\phi)$ are functions of the field $\phi(t)$.  Note that the additive noise case is recovered for $G(\phi)~=~1$.  The first step in obtaining the effective potential for Eq.~(\ref{eq:General_SDE_multiplicative}) is to gain control over noise averaging.  To do so, it is possible to use delta-functionals to represent the formal solution of Eq.~(\ref{eq:General_SDE_multiplicative}) in terms of a constrained path integral:
\begin{eqnarray}
\label{eq:Formal_solution_SDE}
\phi_{\rm sol}(t|\eta) & = & \int {\cal D}\phi \;\phi \;\delta[\phi - \phi_{\rm sol}], \\
                       & = & \int {\cal D}\phi \;\phi \;\delta\left[\frac{d\phi}{dt} + F(\phi) - G(\phi)\eta\right]\sqrt{{\cal J}{\cal J}^{\dagger}}, \;\;\;\;\;\;\;
\end{eqnarray}
where the Jacobian ${\cal J}(\phi)$ is given by:
\begin{eqnarray}
\label{eq:Jacobian_factor}
{\cal J}(\phi) & = & \det\left(\frac{d}{dt} + F'(\phi) - G'(\phi)\eta \right),
\end{eqnarray}
with primes denoting functional derivatives with respect to the field~$\phi$ and ${\cal J}^{\dagger}$ corresponds to the Hermitian conjugate of the operator ${\cal J}$.  

For reasons that will become clear shortly, we define the following functional:
\begin{eqnarray}
\label{eq:Choice_f}
f[\phi_{\rm sol},J] & = & \exp\left(\int dt\; J(t)\phi_{\rm sol}(t)\right).
\end{eqnarray}
where $J(t)$ is an arbitrary (source) function.  Note that here and in the following, functionals are denoted by square brackets $[\;]$).  By (functionally) Taylor expanding Eq.~(\ref{eq:Choice_f}) and substituting Eq.~(\ref{eq:Formal_solution_SDE}), we can show that:
\begin{eqnarray}
\label{eq:Function_phi_solution}
\lefteqn{f[\phi_{\rm sol},J]} \nonumber \\
                       & = & \int {\cal D}\phi \;f[\phi,J] \;\delta\left[\frac{d\phi}{dt} + F(\phi) - G(\phi)\eta\right]\sqrt{{\cal J}{\cal J}^{\dagger}}, \;\;\;\;\;\;\;
\end{eqnarray}
Note that the solution $\phi_{\rm sol}(t|\eta)$ (and thus $f[\phi_{\rm sol},J]$) depends on the noise $\eta$.  Given that the noise has a distribution $P[\eta]$ (assumed to be normalized to unity), it is possible to average the functional $f[\phi,J]$ over all realizations of the noise:
\begin{eqnarray}
\langle f[\phi_{\rm sol},J] \rangle & = & \int{\cal D}\eta\; P[\eta] f[\phi_{\rm sol},J] \\
                                  & = &  \int{\cal D}\eta\; P[\eta]  \int{\cal D}\phi\; f[\phi,J] \nonumber \\
                                  &   &  \times \delta\left[\frac{d\phi}{dt} + F(\phi) - G(\phi)\eta\right]\sqrt{{\cal J}{\cal J}^{\dagger}},
\end{eqnarray}
where we have inserted Eq.~(\ref{eq:Function_phi_solution}) in the last step.   At this stage, we note that the integration over the random fluctuations involves both the delta functional constraint and the Jacobian factor~(\ref{eq:Jacobian_factor}), as the latter also depends on the noise source $\eta$ (this is an important difference with the additive noise case, see Ref.~\cite{Hochberg_etal_1999}).  To proceed with the integration over noise, we use the identity:
\begin{eqnarray}
\lefteqn{\delta\left[\frac{d\phi}{dt} + F(\phi) - G(\phi)\eta \right]} \nonumber \\
  & = & \frac{1}{\det G} \;\delta\left[\eta - \frac{1}{G(\phi)}\left(\frac{d\phi}{dt} + F(\phi)\right) \right],
\end{eqnarray}
Using the above delta functional to integrate over the noise, we obtain:
\begin{eqnarray}
\label{eq:Average_over_noise_f}
\lefteqn{\langle f[\phi_{\rm sol},J] \rangle = \int{\cal D}\phi\; f[\phi,J]} \nonumber \\
  &  &  \times P\left[\frac{1}{G(\phi)}\left(\frac{d\phi}{dt} + F(\phi)\right)\right] \sqrt{{\cal \tilde{J}}{\cal \tilde{J}}^{\dagger}},
\end{eqnarray}
where the Jacobian (solely written in terms of the field $\phi$) is:
\begin{equation}
\label{eq:Jacobian_tilde}
{\cal \tilde{J}}(\phi)  =  \det\left(\frac{1}{G(\phi)}\frac{d}{dt} + \left(\frac{F(\phi)}{G(\phi)}\right)' - \frac{G'(\phi)}{G^{2}(\phi)}\frac{d\phi}{dt} \right).
\end{equation}
Equation~(\ref{eq:Average_over_noise_f}) can be used to average the functional~(\ref{eq:Choice_f}) over the noise.

The next step is to obtain the generating functional for Eq.~(\ref{eq:General_SDE_multiplicative}).  The generating functional contains all the information of a field theory, and allows to compute all correlation functions.  The effective action $\Gamma[\phi]$ (and consequently the effective potential ${\cal V}(\phi)$) can thus be obtained from the generating functional.  The generating functional also connects probability distribution functions and the effective potential (see Ref.~\cite{Hochberg_etal_1999} for details).

To obtain the generating functional, we first assume that the noise is Gaussian
, with zero mean and second moment given by:
\begin{eqnarray}
\label{eq:Noise_shape_function}
N_{\eta}(t,t') & \equiv & \langle \eta(t)\eta(t') \rangle \;=\; A g(t,t'),
\end{eqnarray}
where $A$ is the amplitude of the noise and $g(t,t')$ is a shape function.  We further assume that the noise is time-translation invariant: $g(t,t') = g(t-t')$.  The distribution function for a Gaussian noise can be expressed in the following way:
\begin{equation}
\label{eq:Gaussian_distribution_noise}
P[\eta]  =  C \exp{\left[-\frac{1}{2}\int dt\int dt'\; \eta(t) N_{\eta}^{-1}(t,t') \eta(t')\right]}.
\end{equation}
where $C$ is a normalization constant.  Second, following Ref.~\cite{Hochberg_etal_1999}, we define the generating functional of all correlation functions in the presence of a source $J$ as:
\begin{eqnarray}
Z[J] & \equiv & \langle f[\phi,J] \rangle = \left\langle \exp\left(\int dt\; J(t)\phi(t)\right) \right\rangle
\end{eqnarray}
The motivation for the above definition comes from quantum field theory (e.g. Ref.~\cite{Peskin_Schroeder_1995}).   Substituting Eqs.~(\ref{eq:Gaussian_distribution_noise}) and~(\ref{eq:Choice_f}) into Eq.~(\ref{eq:Average_over_noise_f}), one sees that the average has the same form as the generating functional $Z[J]$ with source $J$ of field theory: 
\begin{eqnarray}
Z[J] & = & \int{\cal D}\phi\; e^{\int dt\; J(t)\phi(t)} \nonumber \\
     &   & \times P\left[\frac{1}{G(\phi)}\left(\frac{d\phi}{dt} + F(\phi)\right)\right] \sqrt{{\cal \tilde{J}}{\cal \tilde{J}}^{\dagger}},
\end{eqnarray}
with the Gaussian noise given by Eq.~(\ref{eq:Gaussian_distribution_noise}).  This generating functional (or partition function) $Z[J]$ can be used to compute averages, correlation functions, etc, and thus contains all the physics of Eq.~(\ref{eq:General_SDE_multiplicative}).  

%
%

Following the usual procedure (e.g. \cite{Peskin_Schroeder_1995}), one can obtain the effective action from the generating functional:
\begin{eqnarray}
\label{eq:Effective_action}
\Gamma[\varphi,\varphi_{0}] & = & S[\varphi] + \frac{A}{2}\left[\ln\det\left(\frac{\delta^{2}S[\varphi]}{\delta\varphi(t)\delta\varphi(t')}\right) \right. \nonumber \\
                      &   & \left. \frac{}{} - \ln {\cal \tilde{J}}[\varphi] - \ln {\cal \tilde{J}}^{\dagger}[\varphi]  \right] - (\varphi\rightarrow\varphi_{0}),
\end{eqnarray}
where the ``classical'' action is:
\begin{eqnarray}
\label{eq:Classical_action}
S[\varphi] & = & \frac{1}{2} \int dt\int dt' \left[\frac{1}{G(\varphi)}\left(\frac{d\varphi(t)}{dt} + F(\varphi)\right)\right] \nonumber \\
        &   & \times g^{-1}(t,t') \left[\frac{1}{G(\varphi)}\left(\frac{d\varphi(t')}{dt'} + F(\varphi)\right)\right].
\end{eqnarray}
Note that the effective action now depends on the ``classical'' field $\varphi(t)$, and not the field $\phi(t)$ (similar to the situation in thermodynamics when changing variables using Legendre transforms).  The classical field is defined as the solution to the classical equation of motion in the presence of a source $J(t)$, obtained by functional differentiation of the classical action:
\begin{eqnarray}
\label{eq:Classical_EOM}
\left.\frac{\delta S}{\delta \phi}\right|_{\phi=\varphi} & = & J(t)
\end{eqnarray}
The notation $(\varphi \rightarrow \varphi_{0})$ in Eq.~(\ref{eq:Effective_action}) indicates a second term similar to the first with $\varphi$ replaced by $\varphi_{0}$, where $\varphi_{0}$ represents the classical field defined in Eq.~(\ref{eq:Classical_EOM}) when $J(t)=0$.

The effective potential is obtained by specializing to static classical field configurations (i.e. $\varphi(t) = \mbox{constant}$) in the effective action.  Some tedious algebra finally gives (see Appendix~\ref{sec:Effective_potential_from_effective_action} for details):
\begin{eqnarray}
\label{eq:Effective_potential_multiplicative_noise}
\lefteqn{{\cal V}(\varphi,\varphi_{0}) = \frac{1}{2}\left(\frac{F(\varphi)}{G(\varphi)}\right)^{2}\int dt\; g^{-1}(t)} \nonumber \\
                        &   & \hspace{0.5in} + \frac{A}{2}\int \frac{d\omega}{(2\pi)} \ln \left[1 + \frac{\hat{g}(\omega)G(\varphi) F(\varphi) \left(\frac{F(\varphi)}{G(\varphi)}\right)''}{\omega^{2} + G^{2}(\varphi)\left[\left(\frac{F(\varphi)}{G(\varphi)}\right)'\right]^{2}}  \right]  \nonumber \\
                        &   & - \mbox{($\varphi\rightarrow\varphi_{0}$)}.
\end{eqnarray}
 where the term proportional to $\left(\frac{F}{G}\right)^{2}$ is the ``classical'' contribution to the potential (i.e. the term that does not depend on the noise amplitude $A$) and the term proportional to the noise amplitude $A$ is the ``fluctuation'' contribution (i.e. the one-loop correction to the classical potential due to the noise).   Note that due to the specialization to static classical field configurations, the effective potential is now a function.  Equation~(\ref{eq:Effective_potential_multiplicative_noise}) gives the effective potential corresponding to differential equations of the type~(\ref{eq:General_SDE_multiplicative}) subjected to multiplicative Gaussian time-translation invariant noise.  It is one of the main results of this paper,  and we illustrate the use (and comment on the physical significance) of Eq.~(\ref{eq:Effective_potential_multiplicative_noise}) in Sect.~\ref{sec:Effective_potential_methods_toy_model}. 

A few comments are in order here.  First, it is easy to check that the effective potential for the additive noise case (originally derived in Ref.~\cite{Hochberg_etal_1999}) is recovered by setting $G(\varphi) = 1$ in Eq.~(\ref{eq:Effective_potential_multiplicative_noise}), as expected.  Second, note the appearance of the integral of the inverse noise shape function in the ``classical'' term in Eq.~(\ref{eq:Effective_potential_multiplicative_noise}).  Its origin can be traced back to the classical action~(\ref{eq:Classical_action}) specialized to static field configurations.  In the following, we require that the shape function $g(t)$ satisfies:
\begin{eqnarray}
\label{eq:Normalization_shape_function}
\int dt\; g^{-1}(t) & = & 1,
\end{eqnarray}
or in Fourier space $\hat{g}^{-1}(\omega = 0) = 1$.  This requirement serves two purposes.  First, it allows to give a precise meaning to the noise amplitude $A$ and the shape function $g(t)$.  In principle, the decomposition~(\ref{eq:Noise_shape_function}) is arbitrary, so normalizing $g(t)$ fixes the value of $A$ (as discussed in Ref.~\cite{Hochberg_etal_1999})).  The second purpose is to remove any remnant of the noise in the tree-level part of the effective potential.  In ordinary quantum field theory, the tree-level part of the effective potential does not depend on fluctuations.  Similarly, when $A = 0$ (no fluctuations), the SDE~(\ref{eq:General_SDE_multiplicative}) becomes a deterministic differential equation, and thus its potential should also not depend on the shape of the noise.  Thus to complete the analogy with ordinary quantum field theory, we impose the condition~(\ref{eq:Normalization_shape_function}).   Third, Eq.~(\ref{eq:Effective_potential_multiplicative_noise}) represents the first two terms in a loop expansion in the noise amplitude $A$ (in analogy to the $\hbar$ expansion in quantum mechanics), and the only assumptions entering into it is that the noise is Gaussian and satisfies the normalization condition~(\ref{eq:Normalization_shape_function}).  It is possible to perform analytical computations with Eq.~(\ref{eq:Effective_potential_multiplicative_noise}) within perturbation theory, but that does not necessarily mean that $A$ is small; an explicit example of this is provided in Sect.~\ref{sec:Effective_potential_methods_toy_model}. 

The effective potential is a construct that is widely used in particle physics to study symmetry breaking in the presence of quantum fluctuations~\cite{Peskin_Schroeder_1995}.  Its construction and interpretation in the context of stochastic partial differential equations subjected to additive noise has been carefully laid out in Refs.~\cite{Hochberg_etal_1999,Hochberg_etal_2000,Hochberg_etal_2000b,Hochberg_etal_2001}.  Some salient physical interpretations and properties (in the context of SDEs) include: (i) stationary points of the effective action correspond to the stochastic expectation values of the field in the absence of an external current; (ii) the effective potential governs the probability that the time-average of the field takes on specific values (read: space-time average, when spatial dependence is included); (iii) the effective potential can be used to obtain the effective ``equations of motion'' in the presence of noise. This latter result is perhaps the most interesting application from the point of view of SDEs, since it implies that we can calculate exactly how the noise shifts the values of the fixed points (we show that explicitly in a toy model example in Sect.~\ref{sec:Effective_potential_methods_toy_model}).  The effective potential has many applications in the context of SDEs.  See for example Ref.~\cite{Hochberg_etal_2000} for an application to the Kardar-Parisi-Zhang equation, where a hydrodynamical interpretation of the dynamical symmetry breaking is discovered and treated using the one-loop effective potential.

In this paper, we instead use the effective potential~(\ref{eq:Effective_potential_multiplicative_noise}) to renormalize the parameters appearing in certain classes of Langevin equations (see Sect.~\ref{sec:Effective_potential_methods_toy_model} for an explicit example).   Said differently: we are not interested in actually solving the said Langevin equations.  The solutions themselves are not relevant to the application of our renormalization objectives.  The constant (stationary) fields, that appear in the argument of our effective potential construct, are the static solutions of the classical equations of motion, obtained from the first variation of the classical action (as discussed in details in Ref.~\cite{Hochberg_etal_1999}).  Our effective potential is expanded about static deterministic classical field configurations, and then the one-loop correction is calculated in perturbation theory. We expand our effective potential around stable stationary deterministic configurations, which guarantees the convexity of our effective potential.  Note that stationarity is defined exactly thus, and {\em not} via stationary probability distributions of the stochastic field (i.e. not via solutions of the Fokker-Planck equation).  We do not require nor need the stochastic field $\phi$ to be stationary, but instead the classical field $\varphi$, as it is defined above.

\section{Application of effective potential methods to a toy model}
\label{sec:Effective_potential_methods_toy_model}

The existence of noise in concert with (non-linear) interactions leads to temporal (and/or spatial) scale-dependence in the parameters of SDEs. And this scaling, can be accounted for quantitatively by solutions of the corresponding RG equations (see Refs.~\cite{Medina_etal_1989,Gagnon_etal_2015,Gagnon_etal_2017,Gagnon_etal_2018} for examples of computation of running parameters using perturbative RG techniques). Just as in quantum field theory, our one-loop effective potential for stochastic field theories~(\ref{eq:Effective_potential_multiplicative_noise}) can be used to identify and then calculate these renormalization group equations and hence, to obtain the scaling in the SDE’s parameters. The scientific significance is this: namely, the dynamics of a stochastic system is modified as we change the scale at which the phenomena is being observed (probed, measured). Since the dynamics is governed by the stochastic equation of motion (i.e., the Langevin equation), we are interested in seeing how the parameters in the Langevin equation scale. Note that our entire analysis is focused on the differential equation itself, and not on its solutions

To illustrate the use of effective potential methods to renormalize SDEs subjected to multiplicative noise, we consider the following toy model:
\begin{eqnarray}
\label{eq:Toy_model_multiplicative_no_diffusion}
\frac{d\phi}{dt} & = &  -r\phi + \lambda \phi^{2} + \phi^{\frac{1}{2}}\eta,
\end{eqnarray}
where $r$ and $\lambda$ are parameters and $\eta(t)$ is a noise term.    The toy model is the same as the one presented in Sect.~\ref{sec:Perturbative_RG}, except that the noise term multiplies $\phi^{\frac{1}{2}}$ instead of $\phi$.  In chemistry, fractional exponents in rate laws could represent the presence of intermediate chemical steps happening at rates that are faster compared to the ``slow'' reaction $2\phi$~+~U~$\rightarrow$~$3\phi$.  Assuming that some of those intermediate chemical steps are light sensitive, then one could subject the chemical reaction to noisy light and measure its effects on the dynamics.  Comparing these measurements to the predictions from the renormalization group, it might be possible to gain insight into those intermediate steps  (see Ref.~\cite{Gagnon_PerezMercader_2017} for details on how the renormalization group can be used to find clues about internal mechanisms of chemical reactions). 

Again for the purpose of illustration, we choose the noise to be Gaussian with second order moment given by Eq.~(\ref{eq:Noise_shape_function}), or in Fourier space:
\begin{eqnarray}
\langle \hat{\eta}(\omega)\hat{\eta}(\omega') \rangle & = & A \hat{g}(\omega) \delta(\omega + \omega'),
\end{eqnarray}
with:
\begin{eqnarray}
\label{eq:Noise_shape_function_toy_model}
\hat{g}(\omega) & = & 1 + \frac{|\omega|}{\omega_{0}},
\end{eqnarray}
where $\omega_{0}$ is some reference frequency scale.  It can be checked that the noise shape function~(\ref{eq:Noise_shape_function_toy_model}) satisfies the condition~(\ref{eq:Normalization_shape_function}).  This Gaussian power-law colored noise has short time correlations, and is the simplest noise that leads to ultraviolet (UV) divergences in the effective potential (as we show explicitly below).  These UV divergences require renormalization, and lead to the running of parameters (such as the removal rate $r$, see below).  Note that white noise does not lead to any UV divergences, and thus cannot be used as an illustration of our effective potential method.

To renormalize this toy model, we first obtain the effective potential using Eq.~(\ref{eq:Effective_potential_multiplicative_noise}), with $F(\phi) = r\phi - \lambda\phi^{2}$ and $G(\phi) = \phi^{\frac{1}{2}}$.  The result is:
\begin{eqnarray}
\label{eq:Effective_potential_toy_model}
{\cal V}(\varphi,\varphi_{0}) & = & \frac{1}{2}\left(r^{2}\varphi -2r\lambda\varphi^{2} + \lambda^{2}\varphi^{3} \right) \nonumber \\
                        &   & + \frac{A}{2}\int \frac{d\omega}{(2\pi)} \ln \left[1 + \frac{\hat{g}(\omega) \left(\frac{-3\lambda}{4}\right) \left(r\varphi - \lambda \varphi^{2}\right)}{\omega^{2} + \frac{r^{2}}{4} - \frac{3r\lambda}{2}\varphi + \frac{9\lambda^{2}}{4}\varphi^{2}}  \right] \nonumber \\
                        &   & - \mbox{($\varphi\rightarrow\varphi_{0}$)}.
\end{eqnarray}
The first term is the ``classical'' potential, and the second term (proportional to $A$) is the noise correction to the classical potential.  One way of renormalizing the toy model~(\ref{eq:Toy_model_multiplicative_no_diffusion}) is to first expand the logarithm in the effective potential~(\ref{eq:Effective_potential_toy_model}):
\begin{eqnarray}
\label{eq:Effective_potential_toy_model_modif1}
{\cal V}(\varphi,\varphi_{0}) & \approx & \frac{1}{2}\left(r^{2}\varphi -2r\lambda\varphi^{2} + \lambda^{2}\varphi^{3} \right) \nonumber \\
                        &   & + \frac{A}{2}\int \frac{d\omega}{(2\pi)} \sum_{n = 1}^{\infty} \frac{(-1)^{n+1}}{n} \nonumber \\
                        &   & \times \left[\frac{\hat{g}(\omega) \left(\frac{-3\lambda}{4}\right) \left(r\varphi - \lambda \varphi^{2}\right)}{\omega^{2} + \frac{r^{2}}{4} - \frac{3r\lambda}{2}\varphi + \frac{9\lambda^{2}}{4}\varphi^{2}}   \right]^{n} \nonumber \\
                        &   & - \mbox{($\varphi\rightarrow\varphi_{0}$)}.
\end{eqnarray}
Since $\hat{g}(\omega) \sim |\omega|$, we see that only the $n=1$ term in the above expansion is UV divergent  (i.e. it is the only term for which the integral is divergent for large frequencies).  Recall that UV divergences are due to short time correlations, and the program of renormalization is to absorb these UV divergences into the parameters of the model.  This ``sweeping-under-the-rug'' of these divergences leads to the running of those parameters with scale (e.g. \cite{Peskin_Schroeder_1995}).  Keeping only the divergent term, we have:
\begin{eqnarray}
\label{eq:Effective_potential_toy_model_modif2}
{\cal V}(\varphi,\varphi_{0}) & \approx & \frac{1}{2}\left(r^{2}\varphi -2r\lambda\varphi^{2} + \lambda^{2}\varphi^{3} \right) \nonumber \\
                        &   & + \frac{3\lambda A}{8}\int \frac{d\omega}{(2\pi)}  \left[\frac{\hat{g}(\omega) \left(-r\varphi + \lambda \varphi^{2}\right)}{\omega^{2} + \frac{r^{2}}{4} - \frac{3r\lambda}{2}\varphi + \frac{9\lambda^{2}}{4}\varphi^{2}}   \right] \nonumber \\
                        &   & - \mbox{($\varphi\rightarrow\varphi_{0}$)}.
\end{eqnarray}
The field-dependent terms in the denominator correspond to infinite resummations of self-energy diagrams in the propagators (as expected from one-loop effective potentials, see for example~\cite{Peskin_Schroeder_1995}).  Expanding the propagator, we obtain:
\begin{eqnarray}
\label{eq:Effective_potential_toy_model_modif3}
{\cal V}(\varphi,\varphi_{0}) & \approx & \frac{1}{2}\left(r^{2}\varphi -2r\lambda\varphi^{2} + \lambda^{2}\varphi^{3} \right) \nonumber \\
                        &   & + \frac{3\lambda A}{8}\int \frac{d\omega}{(2\pi)}  \left[\frac{\hat{g}(\omega) \left(-r\varphi + \lambda \varphi^{2}\right)}{\omega^{2} + \frac{r^{2}}{4}}   \right] \nonumber \\
                        &   & \times \left[1 - \frac{\left(-\frac{3r\lambda}{2}\varphi + \frac{9\lambda^{2}}{4}\varphi^{2}\right)}{\omega^{2} + \frac{r^{2}}{4}} + \dots  \right] \nonumber \\
                        &   & - \mbox{($\varphi\rightarrow\varphi_{0}$)}.
\end{eqnarray}
Since $\hat{g}(\omega) \sim |\omega|$, we see that only the first term in the second bracket is UV divergent.  Keeping only this divergent term, we  obtain:
\begin{eqnarray}
\label{eq:Effective_potential_final_model2_expanded_4}
{\cal V}(\varphi,\varphi_{0}) & \approx & \frac{1}{2}\left(r^{2}\varphi -2r\lambda\varphi^{2} + \lambda^{2}\varphi^{3} \right) \nonumber \\
                        &   & + \frac{3\lambda A}{8}\int \frac{d\omega}{(2\pi)}  \left[\frac{\hat{g}(\omega) \left(-r\varphi + \lambda \varphi^{2}\right)}{\omega^{2} + \frac{r^{2}}{4}}   \right] \nonumber \\
                        &   & - \mbox{($\varphi\rightarrow\varphi_{0}$)}.
\end{eqnarray}
To obtain the corrections to the parameters, we look at the coefficients in front of each power of the field.  The coefficient in front of $\varphi$ is:
\begin{equation}
\label{eq:Correction_1}
\frac{1}{2}r^{2} - \frac{3\lambda A r}{8} \int \frac{d\omega}{(2\pi)}  \;\frac{\hat{g}(\omega)}{\omega^{2} + \frac{r^{2}}{4}},
\end{equation}
and the one in front of $\varphi^{2}$ is:
\begin{eqnarray}
\label{eq:Correction_2}
-\frac{1}{2}(2r\lambda) + \frac{3\lambda^{2} A}{8} \int \frac{d\omega}{(2\pi)}  \;\frac{\hat{g}(\omega)}{\omega^{2} + \frac{r^{2}}{4}}.
\end{eqnarray}
From corrections~(\ref{eq:Correction_1}) and~(\ref{eq:Correction_2}), we infer that only the removal rate $r$ gets a correction at one-loop.  The bare removal rate $r_{0}$ can thus be written as (e.g. \cite{Peskin_Schroeder_1995}):
\begin{eqnarray}
r_{0} & = & r + C \;=\; r\left(1 + \frac{C}{r}\right) \;\equiv\; Z_{r}r,
\end{eqnarray}
where $C$ is the minimal counterterm necessary to cancel the divergence:
\begin{eqnarray}
C & = & -\frac{3\lambda A}{8\omega_{0}} \int \frac{d\omega}{(2\pi)}  \;\frac{|\omega|}{\omega^{2} + \frac{r^{2}}{4}}.
\end{eqnarray}
Regulating the integral using dimensional regularization (e.g. Ref.~\cite{Peskin_Schroeder_1995}) and performing the integration, we get:
\begin{eqnarray}
C & = & -\frac{3\lambda A^{(z)}}{8\omega_{0}} \int \frac{d^{z}\omega}{(2\pi)^{z}}  \;\frac{|\omega|}{\omega^{2} + \frac{r^{2}}{4}} \nonumber \\
          & = & -\frac{3\lambda A^{(z)}}{8\omega_{0}} \; \frac{1}{(4\pi)^{\frac{z}{2}}}\left(\frac{r^{2}}{4}\right)^{\frac{z}{2}-\frac{1}{2}} \frac{\pi}{\Gamma\left(\frac{z}{2}\right) \sin\pi\left(\frac{z}{2}+\frac{1}{2}\right)}, \nonumber \\
\end{eqnarray}
where $A^{(z)}$ is to indicate that the noise amplitude has engineering dimensions that depend on the analytically continued time dimension.  Expanding around the logarithmic pole using $z = 1 - \epsilon$, we get:
\begin{eqnarray}
C  & = & -\frac{3\lambda A^{(z)}}{8\pi\omega_{0}} \; \frac{1}{\epsilon} + \mbox{finite},
\end{eqnarray}
where ``finite'' means terms that are finite as $\epsilon\rightarrow 0$ and which can be dropped for the purpose of finding the running of $r$.  Using the above, the $Z_{r}$ factor can be written as:
\begin{eqnarray}
Z_{r} & = &  1 - \frac{3h^{(z)}}{8\pi} \; \frac{1}{\epsilon},
\end{eqnarray}
with the effective dimensionless rate:
\begin{eqnarray}
\label{eq:Effective_rate}
h^{(z)} & \equiv & \frac{\lambda A^{(z)}}{r\omega_{0}} \;\;=\;\; \frac{\lambda A}{r\omega_{0}}T^{z-1} \;\;=\;\; h\; T^{z-1},
\end{eqnarray}
where $T$ is some arbitrary temporal scale.  To find the running of $h$, we start from:
\begin{eqnarray}
\label{eq:Bare_effective_coupling}
h_{0}^{(z)} & = & \frac{\lambda_{0} A_{0}^{(z)}}{r_{0}\omega_{0}} \;\;=\;\; \frac{\lambda A}{Z_{r} r\omega_{0}} T^{-\epsilon} \;\;=\;\; \frac{1}{Z_{r}} h\; T^{-\epsilon}.
\end{eqnarray}
Taking the derivative with respect to the arbitrary scale $T$ on both sides of Eq.~(\ref{eq:Bare_effective_coupling}), and using the fact that the bare effective rate  $h_{0}$ cannot depend on the arbitrary temporal scale $T$, we obtain the running of the effective rate  with scale:
\begin{eqnarray}
T\frac{dh}{dT} & = & \epsilon h - \frac{3}{8\pi} h^{2}.
\end{eqnarray}
Reverting to the original parameters of the model, we finally obtain the running of the removal rate $r$ at one-loop:
\begin{eqnarray}
\label{eq:Beta_function_decay_rate}
T\frac{dr}{dT} & = & -\epsilon r + \frac{3}{8\pi} \frac{\lambda A}{\omega_{0}}.
\end{eqnarray}
The solution to Eq.~(\ref{eq:Beta_function_decay_rate}) in the $\epsilon\rightarrow 0$ limit is:
\begin{equation}
\label{eq:Running_decay_rate}
r(T) =  r(T^{*}) + \frac{3}{8\pi}\frac{\lambda A}{\omega_{0}}\ln\left(\frac{T}{T^{*}}\right),
\end{equation}
where $T^{*}$ is some time scale at which $r(T^{*})$ is known and can be measured.  This result is valid when  perturbation theory is valid, i.e. when the effective dimensionless rate $h$ is smaller than one (see Eq.~(\ref{eq:Effective_rate})).  Note that this not necessarily imply that the noise amplitude is small, but that the combination of parameters in $h$ is small.  

The physical significance of Eq.~(\ref{eq:Running_decay_rate}) can understood with an analogy with particle physics.  For instance, the properties of an electron (e.g. its electric charge $e$) can be inferred by probing it using a beam of particles with a certain energy.  Since the vacuum surrounding the electron is filled with virtual particles (quantum fluctuations), the electric charge of the electron is screened by this cloud of virtual particles.  Consequently, the measured electric charge depends on how much the beam of particles penetrates into the cloud, and thus on the probing beam energy.  Said differently, the charge of the electron depends on the beam energy $e(E)$, and its variation with energy is given by an equation similar to Eq.~(\ref{eq:Running_decay_rate}).  Since the charge varies with the energy at which it is probed, it can only be defined at a certain reference energy $E^{*}$ at which it can be measured experimentally.  In our example, we use the noise itself as a probe of the chemical reaction represented by Eq.~(\ref{eq:Toy_model_multiplicative_no_diffusion}).   The arbitrary timescale $T$ is akin to the energy $E$ in our particle physics analogy, and represents the coarse-graining scale of the noise.

For fixed values of the noise amplitude $A$, rate of catalysis $\lambda$ and reference frequency $\omega_{0}$, Eq.~(\ref{eq:Running_decay_rate}) shows how the removal rate $r$ varies with the ``probing'' timescale $T$.  The scaling of the system can be seen from Eq.~(\ref{eq:Running_decay_rate}): when the timescale $T$ is re-scaled from say $T_{1}$ to $T_{2}$ and the removal rate from $r(T_{1})$ to $r(T_{2})$, then the system dynamics remains unchanged.  

\begin{figure}
\includegraphics[width=0.45\textwidth]{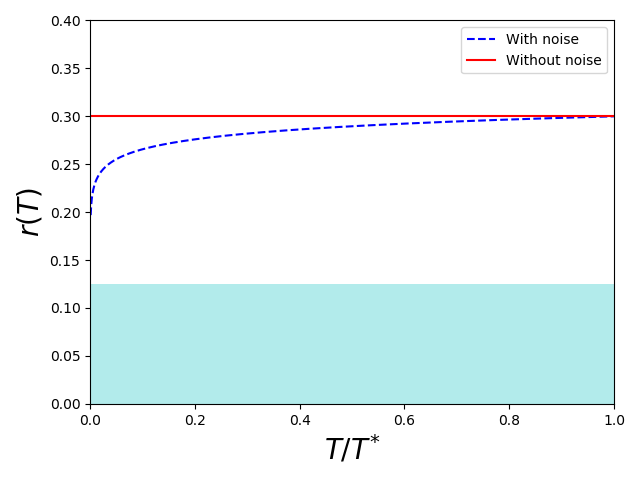}
\caption{Plot of the running removal rate as a function of the scale $T/T^{*}$ (dashed blue line), compared to the situation without noise (red solid line).  We used $A_{v} = 0.5$, $\lambda = 0.05$, $\omega_{0} = 0.2$,  $r(T^{*}) = 0.3$ for the plotting.  Shaded region indicates breakdown of perturbation theory (i.e. $h>1$ for $r < 0.13$).  \label{fig:Plot_running_decay_rate}}
\end{figure}
\begin{figure}
\includegraphics[width=0.45\textwidth]{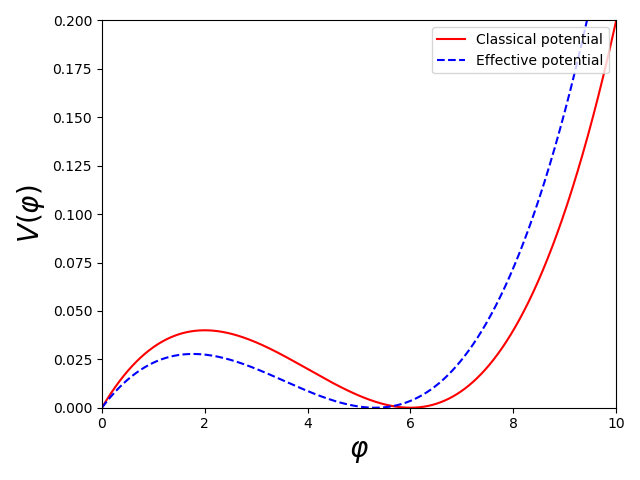}
\caption{Plot of the effective potential as a function of the field $\varphi$ for $T/T^{*} = 0.1$ (dashed blue line), compared to the ``classical'' potential (red solid line).   Note the presence of a shifted minimum (or ``ground state'') for the effective potential.   We used $A_{v} = 0.5$, $\lambda = 0.05$, $\omega_{0} = 0.2$,  $r(T^{*}) = 0.3$ for the plotting.  \label{fig:Plot_effective_potential}}
\end{figure}

A plot of the running of the removal rate is shown in Fig.~\ref{fig:Plot_running_decay_rate}.     The results indicate that noise decreases the removal rate at smaller temporal scales, which constrains possible ``fast'' chemical mechanisms.    The corresponding effective potential can be obtained by substituting the running removal rate~(\ref{eq:Running_decay_rate}) into the ``classical'' potential $\frac{1}{2}\left(\frac{F}{G}\right)^{2}$ (see Eq.~(\ref{eq:Effective_potential_multiplicative_noise})).  A plot of this effective potential is shown in Fig.~\ref{fig:Plot_effective_potential}.  As explained in Sect. VII of Ref.~\cite{Hochberg_etal_1999}, the effective potential governs the probability distribution
of the time average of the fluctuating field $\phi$, and minima of the effective potential correspond to maxima of the probability density of the time-averaged field.  As shown in Ref.~\cite{Hochberg_etal_2000b} for additive noise, a shift in the effective potential's minima implies a shift in the fixed points of the original SDE (without the noise term), and can have important effects on the stability of patterns in reaction-diffusion systems.  In the present toy model example, Fig.~\ref{fig:Plot_running_decay_rate} shows that the removal rate is influenced by the noise, and the shift in the effective potential minimum shown in Fig.~\ref{fig:Plot_effective_potential} indicates the extent by which concentration fixed points are modified by the noise.

Note that the effective potential method used here cannot be used to obtain wavefunction renormalization, due to the restriction to static classical fields when going from Eq.~(\ref{eq:Effective_action}) to Eq.~(\ref{eq:Effective_potential_multiplicative_noise}).  This situation is similar to quantum field theory.  For a full treatment of wavefunction renormalization, it is necessary to analyze the effective action (c.f. Eq.~(\ref{eq:Effective_action})).


\section{Discussion and conclusion}
\label{sec:Conclusion}

A vast range of physical, chemical, and biological phenomena subject to noise and random fluctuations can be investigated by straightforward applications of path integral methods adapted to handle stochastic phenomena.  Provided the phenomenon under study admits a mathematical modeling via nonlinear SDEs subject to additive or multiplicative noise, then a very useful construct known as the effective potential can be defined and explicitly calculated which is intimately related to the PDF of the coarse-grained degrees of freedom.  Note that although we only consider stochastic ordinary differential equations in this paper, the effective potential formalism we developed here for multiplicative noise has been generalized and applied to stochastic partial differential equations subjected to additive noise~\cite{Hochberg_etal_1999,Hochberg_etal_2000,Hochberg_etal_2000b,Hochberg_etal_2001}. 

Normally, out-of-equilibrium SDEs do not admit a description in terms of potentials  (although see the discussion below), but it has been shown in Ref.~\cite{Hochberg_etal_1999,Hochberg_etal_2000,Hochberg_etal_2000b,Hochberg_etal_2001} (for additive noise) and in the present paper (for multiplicative noise) that a parallel can be established with the situation in quantum field theory, and a potential derived that has two major pieces: a ``classical'' contribution plus a ``fluctuation'' contribution.  The classical contribution plus the fluctuation contribution determine the ground state of the system, and therefore the effective potential allows for the determination of the effects of the noise and fluctuations on the ground state of the system, which could be useful in the study of patterns of symmetry, for example.

The fluctuation dependent piece of the potential involves integrations over frequency  (and wavenumber domains, when generalized to partial differential equations).   These integrals require the introduction of a cutoff, which, through renormalization, leads to a scale-dependence of the parameters of the reaction-diffusion equation and therefore has an effect on the type of instability which controls the behavior of the system.   This is illustrated with a toy chemical model in the present paper, where we show explicitly that a specific parameter of the model (i.e. the removal rate) develops a scale dependence when multiplicative power-law noise with short scale correlations is present.  Although the results are obtained in perturbation theory, it shows how the effective potential can be used to incorporate the effects of non-linearities and random fluctuations in the dynamics of out-of-equilibrium systems described by SDEs.


To summarize, the existence of fluctuations (noise) together with interactions leads to scale dependence in the parameters and couplings of stochastic systems.  In which case, the physical/chemical effects of these fluctuations can be accounted for quantitatively by solutions of the corresponding RG equations. These RG equations, which govern the above-mentioned scale dependence in the model parameters, can be conveniently obtained from an effective action and effective potential construct. The explicit solutions of the SDE are not needed in order to set up and calculate (perturbatively, at one loop order) these effective actions or potentials. This is one clear benefit of this approach in so far as we are only interested in uncovering the scale dependence of the model parameters, but not the solutions per se of the SDE. Thus, the purpose of performing these ``formal manipulations'' is to obtain the effective potential, whose divergence structure leads directly to obtaining the desired RG equations. Finally, the solutions of these RG equations yield the scale dependence or running parameters, giving us insight into how the fluctuations together with the (generally nonlinear) interactions induce scale dependence in one or more of the model parameters. Such a scale dependence implies that the value of the parameter depends on the scale of (the temporal and/or spatial) resolution, at which the measurement or observation is made. Hence the dynamics of a (chemical) system evolves as we change the scale at which the (chemical) phenomenon is being observed.

It is important to mention that the effective potential construct discussed here (see Eq.~(\ref{eq:Effective_potential_multiplicative_noise})) is only valid
when condition~(\ref{eq:Normalization_shape_function}) is satisfied and for Gaussian noise (see Eq.~(\ref{eq:Noise_shape_function})).  Note that this does not imply Gaussian fluctuations of the fields, and although this limits the scope of the method, it is sufficient for many practical applications (especially if the noise is external and under the control of the experimentalist).  In addition, the same limitations that apply to the description of phenomena in terms of SDEs also apply to the present method.  For instance, when low number of constituents effects are important (intrinsic noise), an approach based on the Master equation (e.g. \cite{Karzazi_etal_1996,Hansen_etal_2006,Gillespie_2007}) is more appropriate (see also~\cite{Tauber_etal_2005,Cooper_etal_2014} for a mapping between field theory and the Master equation, with applications to the renormalization group).

The notion of effective potential in particle physics is well-developed, and is derived from a field theory viewpoint.  Its generalization to SDEs is subtle, but it has been shown in Ref.~\cite{Hochberg_etal_1999} that the main characteristics of the effective potential that are found useful in particle physics carry over to the SDE case.  On the other hand, there exists other ways of defining a potential for dissipative systems and SDEs~\cite{Graham_Tel_1984,Graham_Tel_1985,Fraikin_Lemarchand_1985}.  These other stochastic potentials have been studied extensively, and it has been shown that they are not analytic in general (e.g. derivatives of the potential diverge near a bifurcation~\cite{Sulpice_etal_1987}).  Whether or not there is a direct connection between stochastic potentials and the effective potential construct presented here, and that it is subject to the same limitations, is interesting and left for future work.

\section*{Acknowledgements}
J.-S. G. and J. P.-M. thank Repsol S. A. for its support and D. H. acknowledges the project CTQ2017-87864-C2-2-P (MINECO) Spain.

\appendix

\section{Obtaining the effective potential from the effective action}
\label{sec:Effective_potential_from_effective_action}

To obtain the final expression for the effective potential~(\ref{eq:Effective_potential_multiplicative_noise}), we start from the effective action~(\ref{eq:Effective_action}):
\begin{eqnarray}
\label{eq:Effective_action_appendix}
\Gamma[\varphi,\varphi_{0}] & = & S[\varphi] + \frac{A}{2}\left[\ln\det\left(\frac{\delta^{2}S[\varphi]}{\delta\varphi(t)\delta\varphi(t')}\right) \right. \nonumber \\
                      &   & \left. \frac{}{} - \ln {\cal \tilde{J}}[\varphi] - \ln {\cal \tilde{J}}^{\dagger}[\varphi]  \right] - (\varphi\rightarrow\varphi_{0}),
\end{eqnarray}
and specialized to constant classical field configurations.  For $\varphi =$ constant,  the classical action becomes:
\begin{eqnarray}
\label{eq:Classical_action_appendix}
S(\varphi) & = & \frac{1}{2} \int dt\int dt' \left[\frac{1}{G(\varphi)}\left(\frac{d\varphi}{dt} + F(\varphi)\right)\right] \nonumber \\
        &   & \times g^{-1}(t,t') \left[\frac{1}{G(\varphi)}\left(\frac{d\varphi}{dt'} + F(\varphi)\right)\right], \nonumber \\
        & = & \frac{1}{2} \int dt\int dt' \left(\frac{F(\varphi)}{G(\varphi)}\right)^{2} g^{-1}(t,t'), \nonumber \\
        & = & \frac{{\cal T}}{2} \left(\frac{F(\varphi)}{G(\varphi)}\right)^{2} \int dt\;  g^{-1}(t),
\end{eqnarray}
where we used Eq.~(\ref{eq:Noise_shape_function}) and the fact that the noise is time-translation invariant in the last step.  The ${\cal T}$ factor represents an infinite time volume that cancels when defining the effective potential (see below).

To obtain the one-loop correction to the classical action, we need to compute the various terms in the square bracket of Eq.~(\ref{eq:Effective_action_appendix}).  Let's start with the first Jacobian term.  For $\varphi =$ constant, the logarithm of the Jacobian term becomes (c.f. Eq.~(\ref{eq:Jacobian_tilde})):
\begin{eqnarray}
\label{eq:Jacobian_factor_appendix}
\ln {\cal \tilde{J}}(\varphi) & = & \ln \det \left[\frac{1}{G}\frac{d}{dt} + \left(\frac{F}{G}\right)'\right],
\end{eqnarray}
where we suppressed the dependence on $\varphi$ for $F(\varphi)$ and $G(\varphi)$ to simplify the notation.  To evaluate the logarithm of the determinant, we use the standard formula (e.g. \cite{Peskin_Schroeder_1995}):
\begin{eqnarray}
\label{eq:Log_det_time_translation_invariant_operator}
\ln\det X(t,t') & = &  {\cal T}\int \frac{d\omega}{(2\pi)}\; \ln \hat{X}(\omega),
\end{eqnarray}
valid for any time-translation invariant operator.  Again the ${\cal T}$ factor represents an infinite time volume.  Using Eq.~(\ref{eq:Log_det_time_translation_invariant_operator}), the logarithm of the Jacobian becomes:
\begin{eqnarray}
\label{eq:Jacobian_factor_appendix_2}
\ln {\cal \tilde{J}}(\varphi) & = & {\cal T}\int \frac{d\omega}{(2\pi)}\; \ln \left[\frac{-i\omega}{G} + \left(\frac{F}{G}\right)' \right].
\end{eqnarray}
The Hermitian conjugate of the Jacobian term is done in a similar way:
\begin{eqnarray}
\label{eq:Jacobian_conjugate_factor_appendix_2}
\ln {\cal \tilde{J}^{\dagger}}(\varphi) & = & {\cal T}\int \frac{d\omega}{(2\pi)}\; \ln \left[\frac{i\omega}{G} + \left(\frac{F}{G}\right)' \right].
\end{eqnarray}
To compute the first term in the square bracket of Eq.~(\ref{eq:Effective_action_appendix}), we need to functionally differentiate the classical action $S[\varphi]$ twice with respect to the classical field $\varphi$, and then specialize to constant classical field configurations.  This gives:
\begin{widetext}
\begin{eqnarray}
\label{eq:Second_variation_classical_action}
\frac{\delta^{2} S[\varphi]}{\delta\varphi(t)\delta\varphi(t')} & = & \frac{G'F}{G^{3}} \int d\tau\; \left[\frac{d\left[g^{-1}(t'-\tau)\right]}{dt'} \right]\delta(t'-t) + \frac{G'F}{G^{3}}\frac{d\left[g^{-1}(t'-t)\right]}{dt'} \nonumber \\
 &  & - \frac{1}{G^{2}}\int d\tau\; \frac{d\left[g^{-1}(t'-\tau)\right]}{dt'}\frac{d\left[\delta(\tau-t)\right]}{d\tau} - \frac{F'}{G^{2}}\frac{d\left[g^{-1}(t'-t)\right]}{dt'} \nonumber \\
 &  & + \frac{F}{G}\left(\frac{F}{G}\right)'' \int d\tau\; \left[g^{-1}(t'-\tau)\right] \delta(t'-t) - \frac{G'F}{G^{2}}\left(\frac{F}{G}\right)' g^{-1}(t'-t) \nonumber \\
 &  & \frac{1}{G}\left(\frac{F}{G}\right)'\int d\tau\; \left[g^{-1}(t'-\tau)\frac{d\left[\delta(\tau-t)\right]}{d\tau}\right] + \frac{F'}{G}\left(\frac{F}{G}\right)' g^{-1}(t'-t),
\end{eqnarray}
where we used the chain rule:
\begin{eqnarray}
\frac{dX[\varphi]}{dt} & = & \frac{\delta X[\varphi]}{d\varphi(t)}\frac{d\varphi(t)}{dt}.
\end{eqnarray}
To compute the logarithm of the determinant of $\delta^{2}S[\varphi]/\delta\varphi(t)\delta\varphi(t')$, we apply formula~(\ref{eq:Log_det_time_translation_invariant_operator}) to Eq.~(\ref{eq:Second_variation_classical_action}), which gives:
\begin{eqnarray}
\label{eq:Log_det_second_variation_action}
\ln\det\left(\frac{\delta^{2} S[\varphi]}{\delta\varphi(t)\delta\varphi(t')}\right) & = & {\cal T} \int\frac{d\omega}{(2\pi)}\;\ln\left[ \frac{G'F}{G^{3}}  (-i\omega)\hat{g}^{-1}(\omega)  + \frac{1}{G^{2}}\omega^{2} \hat{g}^{-1}(\omega) - \frac{F'}{G^{2}}(-i\omega)\hat{g}^{-1}(\omega) + \frac{F}{G}\left(\frac{F}{G}\right)'' \hat{g}^{-1}(\omega = 0)  \right. \nonumber \\
    &   & \left. - \frac{G'F}{G^{2}}\left(\frac{F}{G}\right)' \hat{g}^{-1}(\omega) + \frac{1}{G}\left(\frac{F}{G}\right)' (-i\omega) \hat{g}^{-1}(\omega) + \frac{F'}{G}\left(\frac{F}{G}\right)' \hat{g}^{-1}(\omega) \right].
\end{eqnarray}
In analogy with quantum field theory, we define the effective potential as (e.g. \cite{Peskin_Schroeder_1995}):
\begin{eqnarray}
\Gamma[\varphi,\varphi_{0}] & \equiv & {\cal T} {\cal V}(\varphi,\varphi_{0})
\end{eqnarray}
where we specialized to static classical field configurations (that is why the effective potential is now a function instead of a functional), and the factor ${\cal T}$ is equal to the time volume over which the functional integral is taken.  Collecting all pieces (i.e. Eqs.~(\ref{eq:Classical_action_appendix}), (\ref{eq:Jacobian_factor_appendix_2}), (\ref{eq:Jacobian_conjugate_factor_appendix_2}), (\ref{eq:Log_det_second_variation_action})), we can write down the effective potential:
\begin{eqnarray}
{\cal V}(\varphi,\varphi_{0}) & = & \frac{1}{2} \left(\frac{F}{G}\right)^{2} \int dt\;  g^{-1}(t) + \frac{A}{2}\int\frac{d\omega}{(2\pi)}\left[\ln\left(\frac{G'F}{G^{3}}  (-i\omega)\hat{g}^{-1}(\omega)  + \frac{1}{G^{2}}\omega^{2} \hat{g}^{-1}(\omega) - \frac{F'}{G^{2}}(-i\omega)\hat{g}^{-1}(\omega)  \right.\right. \nonumber \\
                              &   & \left. + \frac{F}{G}\left(\frac{F}{G}\right)'' - \frac{G'F}{G^{2}}\left(\frac{F}{G}\right)' \hat{g}^{-1}(\omega) + \frac{1}{G}\left(\frac{F}{G}\right)' (-i\omega) \hat{g}^{-1}(\omega) + \frac{F'}{G}\left(\frac{F}{G}\right)' \hat{g}^{-1}(\omega) \right) \nonumber \\
                              &   & \left. - \ln \left(\frac{-i\omega}{G} + \left(\frac{F}{G}\right)' \right) - \ln \left(\frac{i\omega}{G} + \left(\frac{F}{G}\right)' \right)\right] - (\varphi\rightarrow\varphi_{0}),
\end{eqnarray}
\end{widetext}
where we used the normalization condition~(\ref{eq:Normalization_shape_function}).  Simplification of the above expression gives the effective potential in Eq.~(\ref{eq:Effective_potential_multiplicative_noise}).

\bibliography{bibliography_file}

\end{document}